\documentclass[12pt]{article}
\usepackage{amssymb}
\usepackage{amsmath}
\usepackage{indentfirst}
\def\a{\alpha}
\def\b{\beta}

\def\d{\delta}
\def\e{\epsilon}

\def\g{\gamma}

\def\h{\eta}

\def\la{\lambda}
\def\m{\mu}
\def\n{\nu}

\def\t{\theta}
\def\r{\rho}
\def\s{\sigma}

\def\x{\xi}
\def\z{\zeta}
\def\D{\Delta}

\def\O{\Omega}

\def\ve{\varepsilon}
\def\vf{\varphi}

\def\vk{\varkappa}

\newcommand{\pd}{\partial}
\newcommand{\Di}{\displaystyle}
\def\be{\begin{equation}}
\def\ee{\end{equation}}
\newcommand{\cC}{{\mathcal{C}}}

\begin{document}

\begin{titlepage}
\setcounter{page}{0}
\begin{flushright}
LMP-TPU-15/13\\
\end{flushright}
\vskip 1cm
\begin{center}
{\LARGE\bf Integrable models associated with}
\vskip 0.3 cm
{\LARGE\bf Myers--Perry--AdS--dS black hole in diverse}
\vskip 0.3 cm
{\LARGE\bf dimensions}\\
\vskip 2cm
$
\textrm{\Large Kirill Orekhov\ }
$
\vskip 0.7cm
{\it
Laboratory of Mathematical Physics, Tomsk Polytechnic University, \\
634050 Tomsk, Lenin Ave. 30, Russian Federation} \\
{Email: orekhovka@tpu.ru}

\end{center}
\vskip 1cm
\begin{abstract} \noindent
Myers--Perry--AdS--dS black hole exhibits $SO(2,1)\times U(n)$ symmetry in the near horizon limit in the special case that all rotation parameters are equal.
We consider a massive relativistic particle propagating on such a background and reduce it to superintegrable spherical mechanics with $U(n)$ symmetry.  A complete set of functionally independent  $u(n)$ generators realized in the model is given.
\end{abstract}


\vskip 0.5cm


\end{titlepage}

\renewcommand{\thefootnote}{\arabic{footnote}}
\setcounter{footnote}0

\noindent
{\bf 1.  Introduction}\\

In the last decade there has been a surge of interest in the Myers-Perry black hole in arbitrary dimension and especially in its near horizon limit (see, e.g., a recent review \cite{kl} and references therein). The first reason
to be concerned about the near horizon geometries
is the duality between the near horizon Kerr black hole and a conformal field theory suggested in \cite{Strom} (for a review see \cite{com}). As was proved in \cite{LuMeiPope}, the duality holds true also in higher dimensions and in the presence of a cosmological constant. The second reason is the possibility to build various conformal mechanics models starting from a massive relativistic particle propagating on such backgrounds \cite{a1}--\cite{GNS}. In this regard the Myers--Perry black hole with all rotation parameters being equal to each other is of particular interest because its symmetry is enlarged to the unitary algebra (in direct sum with extra $so(2,1)$ algebra in the near horizon case) which is the largest finite-dimensional symmetry algebra possible. In particular, this gives a clue to the construction of new maximally superintegrable models in \cite{Gal, GalNers,GNS}.

Note that for a generic conformal mechanics one can always split the radial canonical pair from the angular sector by applying a suitable canonical transformation \cite{HakLecht1, HakLecht2}. The dynamics of the angular variables is governed by the Casimir element in the conformal algebra $so(2,1)$.  The latter can be viewed as the Hamiltonian of a reduced spherical mechanics which retains symmetries pertaining to the angular sector of the parent conformal mechanics.

A natural one--parameter extension of the Myers--Perry solution can be obtained by including a cosmological constant into the consideration \cite{GibbLuPagePope}. It is noteworthy that for the special case that all rotation parameters are equal to each other the configuration retains the unitary symmetry and therefore hints at a possibility to construct new superintegrable models associated with it.  The goal of this work is to construct such superintegrable models which provide a one--parameter deformation of those built recently in \cite{GalNers}.  The similarities and differences between the two cases are discussed in detail.

The paper is organized as follows. In Sect. 2 a short overview of the extremal Myers--Perry--AdS--dS black hole in arbitrary dimension is given. In Sect. 3 we consider such a metric in $D=2n+1$ dimensions for the special case that all rotation parameters are equal. The near horizon limit is implemented and the associated conformal mechanics is constructed. Next we perform the reduction of the conformal mechanics to its spherical sector. Sect. 4 contains a similar analysis for $D=2n$. In Sect. 5 we discuss the unitary symmetries of the Hamiltonians constructed earlier and show that in odd dimensions the reduced Hamiltonian can be expressed in terms of the first and the second order Casimir invariants of the unitary algebra. Finally, we give a complete set of functionally independent $u(n)$ generators realized in the spherical mechanics and prove their superintegrability. In the concluding Sect. 6 we summarize our results and discuss possible further developments.

\newpage

\vspace{0.5cm}

\noindent
{\bf 2.  Myers--Perry--AdS--dS metric in arbitrary dimension}\\

\vspace{0.3cm}

The Myers--Perry--AdS--dS metric is a solution of Einstein equations in $D$ dimensions with a cosmological constant $\la$
\be\label{einstein}
 R_{ij} + (D-1)\la\,g_{ij} = 0,
\ee
which describes a black hole rotating in ($n-\e_D$) spatial two--planes, where $\e_D=0$ for odd dimensions ($D=2n+1$) and $\e_D=1$ for even dimensions ($D=2n$). In Boyer-Lindquist coordinates it reads:
\begin{equation}\label{gen_KadS}
 \begin{array}{rl}
 ds^2 = & \Di{W(1-\la r^2)dt^2 - \frac{U}{\D}dr^2 - \frac{2M}{U}\left(dt-\sum_{i=1}^{n-\e_D}\frac{a_i \m_i^2 d\vf_i}{1+\la a_i^2}\right)^2 - \sum_{i=1}^{n}\frac{r^2+a_i^2}{1+\la a_i^2}d\m_i^2 -}\\
 -& \Di{\sum_{i=1}^{n-\e_D}\frac{r^2+a_i^2}{1+\la a_i^2}\m_i^2(d\vf_i-\la a_i dt)^2 - \frac{\la}{W(1-\la r^2)}\left(\sum_{i=1}^n\frac{r^2+a_i^2}{1+\la a_i^2}\m_i d\m_i\right),}
 \end{array}
\end{equation}
where
\be
 \begin{array}{c}
 \Di{\D = r^{\e_D-2}(1-\la r^2)\prod_{i=1}^{n-\e_D}(r^2+a_i^2),\quad U = r^{\e_D}\sum_{i=1}^{n}\frac{\m_i^2}{r^2+a_i^2}\prod_{j=1}^{n-\e_D}(r^2+a_j^2),}\\
 \Di{W = \sum_{i=1}^n\frac{\m_i^2}{1+\la a_i^2}.}
 \end{array}
\ee
Above $M$ is the black hole mass, $a_i$ are the rotation parameters, $\vf_i$ are azimuthal angles. It is assumed that the latitudinal angular variables $\m_i$ parameterize the unit sphere
\be\label{constraint}
 \sum_{i=1}^n\m_i^2 = 1.
\ee
In even-dimensional case the $n$--th rotation parameter is set to zero
\be
a_n = 0.
\ee

In what follows we shall be mainly concerned with the special case for which all the rotation parameters are equal
\be
a_i = a,
\ee
where $i = 1,\dots, n-\e_D$. In particular, this greatly simplifies the metric (\ref{gen_KadS}). Below we shall treat the even--, and odd--dimensional cases separately.

The black hole solution with equal rotation parameters has a larger symmetry group as one can rotate various spatial two--planes
one into another. In odd dimensions, where the metric takes the form
\begin{equation}\label{KadS_odd}
 \begin{array}{rl}
 ds^2 = & \Di{W(1-\la r^2)dt^2 - \frac{U}{\D}dr^2 - \frac{2M}{U}\left(dt-\frac{a}{1+\la a^2}\sum_{i=1}^n\m_i^2 d\vf_i\right)^2 -}\\
 -& \Di{\frac{r^2+a^2}{1+\la a^2}\sum_{i=1}^n\m_i^2(\la adt-d\vf_i)^2 - \frac{r^2+a^2}{1+\la a^2}\sum_{i=1}^{n}d\m_i^2,}
 \end{array}
\end{equation}
the vector fields generating these rotations can be written as \cite{VasuStevensPage2}
\be\label{un_gen}
 \r_{ij} = x_i\pd_{y_j} - y_j\pd_{x_i} + x_j\pd_{y_i} - y_i\pd_{x_j}, \quad \x_{ij} = x_i\pd_{x_j} - x_j\pd_{x_i} + y_i\pd{y_j} - y_j\pd{y_i},
\ee
Here we introduced coordinates
\be\label{xy}
 x_i = \m_i\cos\vf_i, \quad y_i = \m_i\sin\vf_i ;\quad
 \Di{\vf_i = \arccos\frac{x_i}{\sqrt{x_i^2 + y_i^2}}, \quad \m_i = \sqrt{x_i^2 + y_i^2}},
\ee
which lead also to an equivalent realization:
\be\label{ij}
 \begin{array}{c}
  \Di{\r_{ij} = \sin\vf_{ij}(\m_j\pd_{\m_i} - \m_i\pd_{\m_j}) + \cos\vf_{ij}\left(\frac{\m_j}{\m_i}\pd_{\vf_i} + \frac{\m_i}{\m_j}\pd_{\vf_j}\right),}\\
 \Di{\x_{ij} = -\cos\vf_{ij}(\m_j\pd_{\m_i} - \m_i\pd_{\m_j}) + \sin\vf_{ij}\left(\frac{\m_j}{\m_i}\pd_{\vf_i} + \frac{\m_i}{\m_j}\pd_{\vf_j}\right)},
 \end{array}
\ee
where we denoted $\vf_{ij} = \vf_i - \vf_j$. $\Di{\frac{n(n+1)}{2}}$ generators $\r_{ij}$ and $\Di{\frac{n(n-1)}{2}}$ generators $\x_{ij}$ all together form the unitary algebra $u(n)$.
Note that the existence of the unitary symmetry can be revealed by introducing the complex coordinates
\be
 z_k = \m_k e^{i\vf_k} = x_k + i y_k.
\ee

In even dimensions the metric is
\be
 \begin{array}{rl}
 ds^2 =& \Di{W(1-\la r^2)dt^2 - \frac{U}{\D}dr^2 - \frac{2M}{U}\left(dt-\frac{a}{1+\la a^2}\sum_{i=1}^{n-1}\m_i^2 d\vf_i\right)^2 -}\\
 -& \Di{\frac{r^2+a^2}{1+\la a^2}\sum_{i=1}^{n-1}\m_i^2(d\vf_i-\la adt)^2 - \frac{r^2+a^2}{1+\la a^2}\sum_{i=1}^{n-1}d\m_i^2 - r^2 d\m_n^2 -}\\
 -&\Di{\frac{\la}{W(1-\la r^2)}\left(\frac{a^2(1-\la r^2)}{1+\la a^2}\m_n d\m_n\right)^2},
 \end{array}
\ee
and the angular sector splits into $(\m_i, \vf_i)$, $i=1,\dots ,n-1$ part and $\m_n$ part. After passing to $n-1$ latitudinal angles $\n_i$ (see Sect. 4) the unitary symmetry of $(\n_i, \vf_i)$ sector can be described in exactly the same way as in the odd dimensional case. Therefore in even dimensions metric is invariant under the unitary group $u(n-1)$.

We will also need the expression for the inverse metric which was obtained in \cite{VasuStevensPage2}:
\be\label{gen_inverse}
 \begin{array}{rl}
  g^{\m\n} = & \Di{\left(Q+\frac{(2M)^2}{U\D}\frac{1}{(1-\la r^2)^2}\right)\pd_t^2 - \frac{\D}{U}\pd_r^2 +}\\
  +& \Di{\sum_{i=1}^{n-\e_D}\left(\la aQ + \frac{(2M)^2}{U\D}\frac{1+\la a^2}{(1-\la r^2)^2(r^2+a^2)}+\frac{2M}{U}\frac{a}{(1-\la r^2)(r^2+a^2)}\right)\pd_t\pd_{\vf_i} -} \\
  -&\Di{\sum_{i,j=1}^{n-\e_D}\left(\frac{1+\la a^2}{r^2+a^2}\frac{\d^{ij}}{\m_i^2}-\la^2 a^2 Q+\frac{(2M)^2}{U\D}\frac{a^2(1+\la a^2)^2}{(1-\la r^2)^2(r^2+a^2)^2}-R\right)\pd_{\vf_i}\pd_{\vf_j}+\dots}
  \end{array}
\ee
where the dots denote terms in the $\m_i$--sector which has to be calculated separately for the even--, and odd--dimensional cases. $Q$ and $R$ in (\ref{gen_inverse}) are defined as follows:
\begin{equation}\label{qr}
 \begin{array}{rl}
 Q = & \Di{\frac{1}{W(1-\la r^2)}+\frac{2M}{U}\frac{1}{(1-\la r^2)^2},}\\
 R = & \Di{\frac{(2M)^2}{U\D}\frac{2\la a^2(1+\la a^2)}{(1-\la r^2)^2(r^2+a^2)} + \frac{2M}{U}\frac{a}{(r^2+a^2)^2} +}\\
  +&\Di{\frac{2M}{U}\frac{2\la a^2}{(1-\la r^2)(r^2+a^2)} + \frac{(2M)^2}{U\D}\frac{2a^2(1+\la a^2)}{(1-\la r^2)(r^2+a^2)^2}.}
  \end{array}
\end{equation}
The inverse metric allows one to construct the Hamiltonian of a massive relativistic particle moving on the Myers--Perry--AdS--dS background as a solution $p_0$ of the mass--shell equation $g^{\m\n}p_{\m}p_{\n} = m^2$.

\vspace{0.5cm}

\noindent
{\bf 3.  Odd--dimensional case}\\

\noindent
{\it 3.1  $D=2n+1$ extremal Myers--Perry--AdS--dS black hole near the horizon}\\

For $D = 2n+1$ and equal rotation parameters the metric (\ref{gen_KadS}) can be brought to the form
\begin{equation}\label{KadS1}
 \begin{array}{rl}
 ds^2 = & \Di{\frac{\D}{U}\left(dt-\frac{a}{1+\la a^2}\sum_{i=1}^n\m_i d\vf_i\right)^2 - \frac{U}{\D}dr^2 - \frac{r^2+a^2}{1+\la a^2}\sum_{i=1}^n d\m_i^2 -}\\
 -& \Di{\frac{1}{r^2}\sum_{i=1}^n\m_i^2\left(adt-\frac{r^2+a^2}{1+\la a^2}d\vf_i\right)^2 + \frac{a^2(1-\la r^2)(r^2+a^2)}{r^2(1+\la a^2)^2}\sum_{i<j}^n\m_i^2\m_j^2(d\vf_i-d\vf_j)^2,}
 \end{array}
\end{equation}
where
\be
 \quad U = (r^2+a^2)^{n-1}, \quad \D=\frac{1}{r^2}(1-\la r^2)(r^2+a^2)^n-2M.
\ee

In the extremal case $\D$ has double zero at the horizon radius $r_0$, i.e.:
\begin{equation}
 \D(r_0) = \D^{\prime}(r_0) = 0.
\end{equation}
Solving this equations one can relate the mass and the rotation parameter to the horizon radius $r_0$ and a cosmological constant
\begin{equation}
 a^2 = (n(1-\vk)-1)r_0^2, \quad 2M = \frac{(n r_0^2)^n(1-\vk)^{n+1}}{r_0^2}; \quad \vk := \la r_0^2.
\end{equation}
If one approaches the horizon, i.e. $r \rightarrow r_0 + \ve r_0 r$ followed by $\ve \to 0$, the relations
\begin{equation} \D \rightarrow \ve^2 r_0^2 r^2 V, \quad V := \frac{2(nr_0^2)^{n-1}(1-\vk)^{n-1}(n(1-2\vk)-1)}{r_0^2} \end{equation}
hold.

In order to describe the near horizon geometry, we follow the procedure in \cite{LuMeiPope}. First one makes the coordinate transformation:
\begin{equation}\label{transf1}
 r \rightarrow r_0 + \ve r_0 r, \quad t \rightarrow \frac{\a t}{\ve}, \quad \vf_i \rightarrow \vf_i + \frac{\b_i t}{\ve},
\end{equation}
and then takes the limit $\ve \rightarrow 0$. The number coefficients $\a$ and $\b_i$ above are fixed from the condition that the first two terms in (\ref{KadS1}) produce the $AdS_2$ metric up to a factor, while the rest is nonsingular
\begin{equation}\label{ab1}
 \a = \frac{r_0^2+a^2}{2r_0(n(1-2\vk)-1)}, \quad \b_i = \frac{a(1+\la a^2)}{2r_0(n(1-2\vk)-1)}.
\end{equation}

The near horizon extremal metric reads
\be\label{ex_KadS1}
 \begin{array}{rl}
 ds^2 = & \Di{\frac{r_0^2}{2(n(1-2\vk)-1)}\left(r^2 dt^2-\frac{dr^2}{r^2}\right) -\frac{a^2}{(n(1-2\vk)-1)^2}\sum_{i=1}^n\m_i^2(rdt+d\vf_i)^2-}\\
  -&\Di{\frac{r_0^2+a^2}{1+\la a^2}\sum_{i=1}^n d\m_i^2 + \frac{a^4}{nr_0^2(n(1-2\vk)-1)^2}\sum_{i<j}^n\m_i^2\m_j^2(d\vf_i-d\vf_j)^2,}
  \end{array}
\ee
where we rescaled the azimuthal angular variables as follows:
\be\label{transf11}
\vf_i \rightarrow \frac{ar_0(1+\la a^2)}{(r_0^2+a^2)(n(1-2\vk)-1)}\vf_i.
\ee
It is readily verified that (\ref{ex_KadS1}) is a vacuum solution of the Einstein equations with a cosmological constant (\ref{einstein}). It is an extension of the metric constructed in \cite{Gal} which now includes a cosmological constant $\la$.

A salient feature of the near horizon metric (\ref{ex_KadS1}) is that it exhibits extra symmetries generated by the Killing vectors
\be\label{killing1}
D = t\,\pd_t-r\,\pd_r, \quad K = \left(t^2+\frac{1}{r^2}\right)\pd_t - 2tr\,\pd_r -\frac{2}{r}\sum_{i=1}^n\pd_{\vf_i},
\ee
which along with the time translation $H = \pd_t$ form the conformal algebra $so(2,1)$.
\vspace{0.5cm}

{\it 3.2 Conformal mechanics near the horizon of the extremal Myers--Perry--AdS--dS black hole in $D=2n+1$}\\

In order to construct the Hamiltonian of a massive relativistic particle moving on the curved background (\ref{ex_KadS1}), we first invert the metric~\footnote{The constant factor of $\Di{\frac{2(n(1-2\vk)-1)}{r_0^2}}$ has been removed by redefining $m^2$. Since the $\m_i$ sector in (\ref{ex_KadS1}) does not mix with other coordinates, the corresponding piece in the metric can be inverted separately.}
\be\label{inverse1}
 \begin{array}{rl}
  g^{\m\n}\pd_{\m}\pd_{\n} &=  \Di{\frac{\pd_t^2}{r^2} - r^2\pd_r^2 + \frac{r_0^2(1+\la a^2)}{2(r_0^2+a^2)(n(1-2\vk)-1)}\sum_{i,j=1}^{n-1}(\m_i\m_j-\d_{ij})\pd_{\m_i}\pd_{\m_j}-}\\
  &+ \Di{\sum_{i,j=1}^n\left(\frac{(1-\vk)(n(1-2\vk)-1)}{2(1+\la a^2)} + 1 -\frac{(r_0^2+a^2)(n(1-2\vk)-1)}{2a^2(1+\la a^2)}\frac{\d_{ij}}{\m_i^2}\right)\pd_{\vf_i}\pd_{\vf_j}-}\\
 &-\Di{\frac{2}{r}\sum_{i=1}^n\pd_t\pd_{\vf_i}},
 \end{array}
\ee
and then solve the mass--shell condition $g^{\m\n}p_{\m}p_{\n} = m^2$
for the energy
\be\label{ham1}
 \begin{array}{c}
 \Di{H = r\left(\sqrt{\O}-\sum_{i=1}^n p_{\vf_i}\right),} \\
 \Di{\O = m^2 + (rp_r)^2 +\h\sum_{i,j=1}^{n-1}(\d_{ij}-\m_i\m_j)p_{\m_i}p_{\m_j} +\sum_{i,j=1}^n\left(\tau\frac{\d_{ij}}{\m_i^2}-\s\right)p_{\vf_i}p_{\vf_j}},\\[12pt]
 \Di{\h=\frac{r_0^2(1+\la a^2)}{2(r_0^2+a^2)(n(1-2\vk)-1)}, \quad \s=\frac{(1-\vk)(n(1-2\vk)-1)}{2(1+\la a^2)},}\\[12pt]
 \Di{\tau=\frac{(r_0^2+a^2)(n(1-2\vk)-1)}{2a^2(1+\la a^2)}.}
 \end{array}
\ee

Associated with the Killing vectors (\ref{killing1}) are the integrals of motion
\be\label{integrals1}
 \begin{array}{c}
 \Di{H = r\left(\sqrt{\O}-\sum_{i=1}^n p_{\vf_i}\right), \quad D = tH+rp_r, \quad K = t^2 H + 2trp_r + \frac{1}{r}\left(\sqrt{\O}+\sum_{i=1}^n p_{\vf_i}\right),}
 \end{array}
\ee
which form $so(2,1)$ algebra under the Poisson bracket:
\be
 \{H,D\} = H, \quad \{H,K\} = 2D, \quad \{D,K\} = K,
\ee
Computing the Casimir invariant of the $so(2,1)$ algebra
\be\label{cas1}
 C = HK-D^2 + P^2= m^2 + \h\sum_{i,j=1}^{n-1}(\d_{ij}-\m_i\m_j)p_{\m_i}p_{\m_j} +\sum_{i,j=1}^n\left(\tau\frac{\d_{ij}}{\m_i^2}-\s\right)p_{\vf_i}p_{\vf_j},
\ee
where we added integral of motion $P^2 = \sum_{i=1}^n(p_{\vf_i})^2$ for convenience, one finds a function on the phase space which depends only on the angular variables and is quadratic in the momenta. Following the ideology in \cite{HakLecht1, HakLecht2}, it can be considered to be the Hamiltonian of a reduced spherical mechanics. By construction, it inherits the $U(n)$--symmetry of the background, while the decoupling of the radial coordinate is achieved at the expanse of missing $SO(2,1)$. The system (\ref{cas1}) is a one--parameter deformation of the model studied in \cite{Gal,GalNers}. The detailed discussion of its unitary symmetry and integrability is given below in Sect. 5.

Note that since the Hamiltonian (\ref{ham1}) does not depend on the azimuthal angular variables $\vf_i$, the momenta $p_{\vf_i}$ are conserved in time. Setting them to be coupling constants yields a further reduction which, up to a redefinition of the coupling constants, coincides with the maximally superintegrable model analyzed in \cite{GalNers}.

\vspace{0.5cm}

\noindent
{\bf 4.  Even-dimensional case}\\

{\it 4.1 $D=2n$ extremal Myers--Perry--AdS--dS black hole near the horizon}\\

For $D=2n$ and equal rotation parameters the metric (\ref{gen_KadS}) can be brought to the form
\be\label{KadS2}
 \begin{array}{rl}
  ds^2=& \Di{\frac{\D}{U}\left(dt-\frac{a}{1+\la a^2}\sum_{i=1}^{n-1}\m_i^2 d\vf_i\right)^2 - \frac{U}{\D}dr^2 - \frac{r^2+a^2}{1+\la a^2}\sin^2\t\sum_{i=1}^{n-1} d\n_i^2 - \frac{\r^2}{\D_{\t}}d\t^2 -}\\
  -& \Di{\frac{\D_{\t}}{\r^2}\sum_{i=1}^{n-1}\m_i^2\left(adt-\frac{r^2+a^2}{1+\la a^2}d\vf_i\right)^2 + \frac{a^2(r^2+a^2)(1-\la r^2)}{(1+\la a^2)^2\D_{\t}}\sum_{i<j}^{n-1}\m_i^2\m_j^2(d\vf_i-d\vf_j)^2},
  \end{array}
\ee
where we introduced one spherical angle $\theta$
\be\label{sph_coor}
 \m_i = \n_i\sin\t \, \quad \m_n = \cos\t, \quad \sum_{i=1}^{n-1}\n_i^2 = 1
\ee
and denoted
\be\begin{array}{c}
 \Di{\quad U = \frac{1}{r}(r^2+a^2\m_n^2)(r^2+a^2)^{n-2}, \quad \D = \frac{1-\la r^2}{r}(r^2+a^2)^{n-1}-2M,}\\
 \Di{\r = r^2+a^2\cos^2\t, \quad \D_{\t} = 1+\la a^2\cos^2\t.}
\end{array}\ee

Imposing the extremality condition
\be
 \D(r_0) = \D^{\prime}(r_0) = 0,
\ee
one can link the black hole mass and a cosmological constant to the horizon radius and the rotation parameter
\be
 \la = \frac{(2n-3)r_0^2-a^2}{r_0^2(a^2+(2n-1)r_0^2)}, \quad M = \frac{(r_0^2+a^2)^n}{r_0(a^2+(2n-1)r_0^2)}.
\ee

In order to construct the near-horizon limit, it suffices to change the coordinates
\be\label{transf2}
 r \rightarrow r_0 + \ve r_0 r, \quad t \rightarrow \frac{\a t}{\ve}, \quad \vf_i \rightarrow \vf_i + \frac{\b_i t}{\ve},
\ee
take $\a$ and $\b_i$ in the form
\be
 \a = \frac{r_0^2+a^2}{r_0 V}, \quad \b_i = \frac{a(1+\la a^2)}{r_0 V}, \quad V = \frac{\tilde{V}}{(r_0^2+a^2)^{n-2}}.
\ee
and finally send $\ve$ to zero. This yields
\be\label{ex_KadS2}
 \begin{array}{rl}
 ds^2 =& \Di{\frac{\r_0^2}{V}\left(r^2 dt^2-\frac{dr^2}{r^2}\right) - \frac{r_0^2+a^2}{1+\la a^2}\sin^2\t\sum d\n_i^2 - \frac{\r_0^2}{\D_{\t}}d\t^2 -}\\
 -& \Di{\frac{\D_{\t}}{\r_0^2}\frac{4a^2 r_0^2}{V^2}\sum\m_i^2(rdt+d\vf_i)^2 + \frac{4a^4 r_0^2(1-\la r_0^2)}{\r_0^2(r_0^2+a^2)V^2}\sum_{i<j}\m_i^2\m_j^2(d\vf_i-d\vf_j)^2},\\
 \r_0^2 = & r_0^2 + a^2\cos^2\t,
 \end{array}
\ee
which is a vacuum solution of the Einstein equations in the presence of a cosmological constant. Note that,
when deriving the last formula, we rescaled the azimuthal angular variables
\be\label{transf21}
 \vf_i \rightarrow \frac{2ar_0}{V}\frac{1+\la a^2}{r_0^2+a^2}\vf_i
\ee
and have taken into account the following relations:
\be
 \D \rightarrow \ve^2 r_0^2 r^2\tilde{V}, \quad \tilde{V} := \frac{(r_0^2+a^2)^{n-2}(a^4+2a^2(2n-1)r_0^2-(3-8n+4n^2)r_0^4)}{r_0^2(a^2+(2n-1)r_0^2)}
\ee
which hold true in the near horizon limit.

\vspace{0.5cm}

{\it 4.2 Conformal mechanics near the horizon of the extremal Myers--Perry--AdS--dS black hole in $D=2n$}\\

Like in the preceding section, we shall construct the Hamiltonian of a conformal mechanics associated with the near horizon geometry of the extremal Myers--Perry--AdS--dS black hole in $D=2n$ by first inverting the metric
\be\label{inverse2}
 \begin{array}{rl}
  g^{\m\n}\pd_{\m}\pd_{\n} =& \Di{\frac{V}{\r_0^2}\left(\frac{\pd_t^2}{r^2}-r^2\pd_r^2\right) - \frac{1+\la a^2}{(r_0^2+a^2)\sin^2\t}\sum_{i,j=1}^{n-2}(\d_{ij}-\n_i\n_j)\pd_{\n_i}\pd_{\n_j} - \frac{\D_{\t}}{\r_0^2}\pd_{\t}^2 -}\\
  -& \Di{\sum_{i,j=1}^{n-1}\left(\frac{V^2(r_0^2+a^2)}{4a^2r_0^2(1+\la a^2)}\frac{\d_{ij}}{\m_i^2} - \frac{V^2(r_0^2+a^2)}{2\D_{\t}r_0^2(a^2+(2n-1)r_0^2)(1+\la a^2)} - \frac{V}{\r_0^2}\right)\pd_{\vf_i}\pd_{\vf_j}-}\\
 -&\Di{2\frac{V}{\r_0^2\, r}\sum_{i=1}^{n-1}\pd_t\pd_{\vf_i}}
  \end{array}
\ee
and then solving the mass--shell condition for the energy
\be\label{ham2}
 \begin{array}{rl}
 H = & \Di{r\left(\sqrt{\O} - \sum_{i=1}^{n-1}p_{\vf_i}\right)}, \\
 \O = & \Di{\frac{m^2\r_0^2}{V}+ (rp_r)^2 +\sum_{i=1}^{n-1}(p_{\vf_i})^2+ \frac{1+\la a^2}{V(r_0^2+a^2)}\frac{\r_0^2}{\sin^2\t}\sum_{i,j=1}^{n-2}(\d_{ij}-\n_{i}\n_{j})p_{\n_i}p_{\n_j} + \frac{\D_{\t}}{V}p_{\t}^2} \\
 + & \Di{\sum_{i,j=1}^{n-1}\left(\frac{V(r_0^2+a^2)\r_0^2}{4a^2 r_0^2(1+\la a^2)}\frac{\d^{ij}}{\m_i^2} - \frac{V(r^2_0+a^2)}{2r_0^2(a^2+(2n-1)r_0^2)(1+\la a^2)}\frac{\r_0^2}{\D_{\t}}-1\right)p_{\vf_i}p_{\vf_j}}.
 \end{array}
\ee
This Hamiltonian possesses conformal symmetry generated by the Killing vectors (\ref{killing1}) which gives rise to the integrals of motion realized as in (\ref{integrals1}) with $\O$ now taken from the previous line.

Computing the Casimir element in the conformal algebra, one gets the Hamiltonian of the spherical mechanics related to the near horizon geometry of the extremal Myers--Perry--AdS--dS black hole in $D=2n$
\be\label{cas2}
 \begin{array}{rl}
 C = & \Di{HK-D^2 + P^2=} \\
 =& \Di{\frac{m^2\r_0^2}{V}+ \h\frac{\r_0^2}{\sin^2\t}\sum_{i,j=1}^{n-2}(\d_{ij}-\n_{i}\n_{j})p_{\n_i}p_{\n_j} + \frac{\D_{\t}}{V}p_{\t}^2 + \r_0^2\sum_{i,j=1}^{n-1}\left(\tau\frac{\d^{ij}}{\m_i^2} - \frac{\s}{\D_{\t}}\right)p_{\vf_i}p_{\vf_j}},\\
\tau = & \Di{\frac{V(r_0^2+a^2)}{4a^2 r_0^2(1+\la a^2)}, \quad \s = \frac{V(r^2_0+a^2)}{2r_0^2(a^2+(2n-1)r_0^2)(1+\la a^2)}, \quad \h = \frac{1+\la a^2}{V(r_0^2+a^2)}},
 \end{array}
\ee
where integral of motion $P^2 = \sum_{i=1}^{n-1}(p_{\vf_i})^2$ was added for convenience. As compared to the model constructed in \cite{Gal,GalNers}, the Hamiltonian (\ref{cas2}) is deformed by the terms which depend on a cosmological constant and, as thus, it provides a one--parameter continuous deformation of the former. The detailed discussion of its unitary symmetry and integrability is given in the next section.

Because the  azimuthal angular variables are cyclic, one can consider a further reduction of (\ref{cas2}) which is obtained by setting the
angular momenta $p_{\vf_i}$ to be coupling constants. This gives the Hamiltonian
\be\label{sp_ham2}
 \Di{\tilde{H} = \frac{m^2\r_0^2}{V}+ \frac{\D_{\t}}{V}p_{\t}^2 - \s^{\prime}\frac{\r_0^2}{\D_{\t}}+ \h^{\prime}\frac{\r_0^2}{\sin^2\t}\left(\sum_{i,j=1}^{n-2}(\d_{ij}-\n_{i}\n_{j})p_{\n_i}p_{\n_j} + \sum_{i,j=1}^{n-1}\frac{\g_i^2}{\n_i^2}\right)},
\ee
$m^2$, $\s^{\prime}$, $\h^{\prime}$ ang $\gamma_i$ are the coupling constants.
This model is a one--parameter deformation of that in \cite{GalNers}.

The proof of superintegrability of (\ref{sp_ham2}) is not affected by the presence of a cosmological constant and proceeds along the same lines as in
\cite{GalNers}. The expression in braces is the maximally superintegrable model studied in \cite{GalNers}.
In this sector one can realize $2(n-2)-1$ functionally independent integrals of motion. The full system (\ref{sp_ham2}) involves one more canonical pair and only one extra integral of motion (the Hamiltonian (\ref{sp_ham2}) itself). The model thus lacks for one integral of motion to be maximally superintegrable.

\vspace{0.5cm}

\noindent
{\bf 5. Unitary symmetry and superintegrability}\\

Let us discuss symmetries and superintegrability of the spherical mechanics constructed above in more detail. Consider first the odd-dimensional case for which the dynamics is governed by the Hamiltonian (\ref{cas1}). By construction, it inherits from the parent Hamiltonian (\ref{ham1}) the $U(n)$--symmetry realized in the angular sector. The corresponding Killing vector fields are given in
(\ref{un_gen}). In particular, one can verify that the Hamiltonian can be expressed via the linear and the quadratic Casimir invariants of $u(n)$
\be
\cC_1 =  \frac{1}{2}\sum_{i=1}^n\r_{ii}, \quad \cC_2 = \frac{1}{2}\sum_{i,j=1}^n(\r_{ij}^2 + \x_{ij}^2)
\ee
as follows
\be\label{sph}
H^{sph}_n = \cC_2 - \cC_1^2= \sum_{i,j=1}^{n-1}(\d_{ij}-\m_i\m_j)p_{\m_i}p_{\m_j} +\sum_{i=1}^n\frac{p_{\vf_i}^2}{\m_i^2}.
\ee
For later convenience we invert the transformation (\ref{transf11}) and drop the arising constant multiple and a constant term in (\ref{cas1}) casting the Hamiltonian into the form
\be\label{cc}
 \begin{array}{rl}
 C \equiv H_n = &  \Di{\tau H^{sph}_n - \s\cC_1^2 =} \\
 = & \tau\cC_2 - (\s + \tau)\cC_1^2.
\end{array}
\ee
This formula shows that $u(n)$ is the spectrum generating algebra of the system. This property is particularly useful in quantum mechanics because a well developed group theoretical framework is available to construct its eigenstates and eigenvalues (see e.g. \cite{Iachello}).

Let us discuss integrability of the system governed by the Hamiltonian $H_n$ which involves ${2n-1}$ configuration space degrees of freedom. There are $n$ first order Casimir invariants $\cC_1(u(1)),\dots ,\cC_1(u(n))$ which together with $n-1$ second order ones $\cC_2(u(2)),\dots ,\cC_2(u(n))$ form a set of $2n - 1$ functionally independent integrals of motion in involution. Therefore this system is Liouville integrable. The issue of superintegrability is more involved because one needs to count the number of functionally independent integrals of motion among $n^2$ generators $\r_{ij}, \x_{ij}$ of $u(n)$.

Let us use the coordinates $(x_i, y_i)$ (\ref{xy}), in which $\r_{ij}$ and $\x_{ij}$ read
\be\label{un_gen1}
 \r_{ij} = x_i p_{y_j} - y_j p_{x_i} + x_j p_{y_i} - y_i p_{x_j}, \quad \x_{ij} = x_i p_{x_j} - x_j p_{x_i} + y_i p_{y_j} - y_j p_{y_i}.
\ee
These expressions provide a canonical realization of $u(n)$. 

The number of functionally independent integrals of motion is equal to the rank of the matrix $\pd_{\z_a}I_b$, $\z_a$ denote all the phase space coordinates and $I_b$ designates the generators. The case $n = 1$ is trivial. There is one configuration space degree of freedom and one integral of motion. For $n = 2$ there are eight coordinates $\z_a$ and four integrals of motion $I_b$. One can verify that $\text{rank}(\pd_{\z_a}I_b) = 4$ meaning that all  $I_b$ are independent. For $n = 3$ there are twelve coordinates and nine integrals of motion. However, in this case $\text{rank}(\pd_{\z_a}I_b) = 8$ which implies that one integral is a function of the others. A relation between them can be written explicitly
\be\label{rel}
 \frac{1}{2}\left(\r_{11}(\r_{23}^2 + \x_{23}^2) + \r_{22}(\r_{13}^2 + \x_{13}^2) + \r_{33}(\r_{12}^2 + \x_{12}^2)\right) = \r_{23}(\x_{12}\x_{13} + \r_{12}\r_{13}) + \x_{23}(\r_{12}\x_{13} - \x_{12}\r_{13}).
\ee
Note that
this relation is of the third order in generators and it does not occur in the completely reduced case because it can not be expressed in terms of $I_{ij} = \r_{ij}^2 + \x_{ij}^2, i<j,$ only (cf. \cite{GalNers}).

We see that for $n = 2$ and $n = 3$ the number of functionally independent integrals of motion is $4n - 4$. This holds true for all $n \geq 2$ which can be proved by induction. Let us assume that for some $N = n - 1$ there are $4 (n-1) - 4$ functionally independent integrals of motion, which one can choose as follows:
\be\label{set}
\r_{11}, ~ \r_{12}, ~ \x_{12}, ~ \r_{22}, ~ \r_{1i}, ~ \x_{1i}, ~ \r_{2i}, ~ \x_{2i},
\ee
where $i=3,\dots ,n-1$. Then for $N = n$ one adds $2n - 1$ integrals $\r_{in}$ and $\x_{in}$ with $i = 1,\dots ,n-1$ as well as $\r_{nn}$. For each pair of the integrals  $\r_{in}$ and  $\x_{in}$, where  $i = 3,\dots ,n-1$, let us consider the following columns in the matrix $\pd_{\z_a}I_b$:
\be
 \pd_{\z_a}\{\r_{11},\r_{1i},\x_{1i},\r_{ii},\r_{1n},\x_{1n},\r_{in},\x_{in},\r_{nn}\}, \quad \pd_{\z_a}\{\r_{2},\r_{2i},\x_{2i},\r_{ii},\r_{2n},\x_{2n},\r_{in},\x_{in},\r_{nn}\}
\ee
These columns have exactly the same structure as for $n = 3$, provided one makes the substitutions of indices $(123) \rightarrow (1in)$ and $(123) \rightarrow (2in)$. Therefore they lead to the same relations between the generators as in (\ref{rel})
\be\label{rel2}
  \r_{kk}(\r_{in}^2 + \x_{in}^2) + \r_{ii}(\r_{kn}^2 + \x_{kn}^2) + \r_{nn}(\r_{ki}^2 + \x_{ki}^2) = \r_{in}(\x_{ki}\x_{kn} + \r_{ki}\r_{kn}) + \x_{in}(\r_{ki}\x_{kn} - \x_{ki}\r_{kn}),
\ee
where $k = 1,2$. In order to determine $\r_{nn}$ as a function of other generators, we consider another set of columns
\be
 \pd_{\z_a}\{\r_{11},\r_{12},\x_{12},\r_{22},\r_{1n},\x_{1n},\r_{2n},\x_{2n},\r_{nn}\},
\ee
which leads to the same relation as in (\ref{rel2}) with $k = 1, i = 2$. We thus conclude that the generators $\r_{1n}, \x_{1n}, \r_{2n}, \x_{2n}$ are functionally independent. Together with (\ref{set}) they form a complete set of $4n - 4$ functionally independent integrals of motion which completes the induction. It follows from the previous discussion that the spherical mechanics in odd dimensions lacks for only one integral of motion to be maximally superintegrable.

The analysis of the even dimensional case with the dynamics governed by the Hamiltonian (\ref{cas2}) proceeds along the same lines. First one inverts the transformation (\ref{transf21}) which brings the Hamiltonian to the form
\be\label{h2}
	 C \equiv \tilde{H}_n = \frac{m^2\r_0^2}{V} + \frac{\D_{\t}}{V}p_{\t}^2 + \tau\r_0^2 H^{sph}_{n-1} - \frac{\s}{\D_{\t}}\left(\sum_{i=1}^{n-1}p_{\vf_i}\right)^2,
\ee
where
\be
 \tau = \frac{1+\la a^2}{V(r_0^2 + a^2)}, \qquad \s = \frac{2a^2 (1+ \la a^2)}{V(r_0^2 + a^2)(a^2 + (2n-1)r_0^2)},
\ee
and $H^{sph}_{n-1}$ is defined in (\ref{sph}). This system has $2n - 2$ configuration space degrees of freedom and its Liouville integrability is ensured by the existence of $2n - 2$ commuting independent integrals of motion $H^{sph}_2,\dots ,H^{sph}_{n-1}, p_{\vf_1},\dots ,p_{\vf_{n-1}}, \tilde{H}_n$. It has the same symmetry algebra as $H^{sph}_{n-1}$, i.e. $u(n-1)$. The complete set of $4n - 7$ functionally independent integrals of motion reads
\be
 \tilde{H}_n, \r_{11}, \r_{12}, \x_{12}, \r_{22}, \r_{1i}, \x_{1i}, \r_{2i}, \x_{2i}, 
\ee
where $i = 3,\dots , n - 1$. Therefore the system lacks for two independent integrals of motion to be maximally superintegrable.

\vspace{0.5cm}

{\bf 6. Conclusion}\\

To summarize, in this work we have constructed mechanical systems with the conformal and unitary symmetry which result from the near horizon Myers--Perry--AdS--dS black hole geometry in arbitrary dimension. We presented both the Hamiltonians and the integrals of motion as well as performed a reduction to a spherical mechanics which is governed by the Casimir invariant of the conformal group $SO(2,1)$. These models provide one-parameter deformations of the systems constructed recently in \cite{Gal, GalNers,GNS}. It was demonstrated that they are superintegrable but not maximally superintegrable, lacking one integral of motion in the odd--dimensional case and two integrals of motion in the even--dimensional case. A canonical realization of the unitary algebra (\ref{un_gen1}) was studied and the functionally independent generators were identified.

A further reduction of these models was attained by setting momenta canonically conjugate to the azimuthal angular variables to be coupling constants. It was shown that, up to a redefinition of constants, the resulting Hamiltonian in odd dimensions is the same as in the case of a vanishing cosmological constant \cite{GalNers}. In even dimensions, however, there are extra terms in the reduced Hamiltonian but their presence does not alter the number of functionally independent integrals of motions.

There are several possible developments of this work. A generalization of the present consideration to the case of nonequal rotational parameters is of considerable interest. The case of non--vanishing electromagnetic field is worthy studying as well. And also, new models can be obtained using contractions of $u(n)$ algebra \textit{a-l\'a} Smorodinsky-Winternitz (see e. g. \cite{w} and the references therein).

\vskip 0.5cm
\noindent
{\bf Acknowledgements}\\

\noindent
This work was supported by the Dynasty Foundation and RFBR grant 13-02-90602-Arm.

\end{document}